\documentclass[preprint,12pt]{elsarticle}
\biboptions{sort&compress}

\usepackage{amssymb}
\usepackage{amsmath}
\usepackage{xurl}
\usepackage{tikz}
\usepackage{amsfonts}
\usepackage{array}
\usepackage[caption=false]{subfig}
\usepackage{makecell}
\usepackage{booktabs}
\usepackage{tablefootnote}

\DeclareMathOperator*{\argmin}{arg\,min}

\newcommand{\rev}[1]{\textcolor{black}{#1}}

\definecolor{seabornblue}{rgb}{0.298, 0.447, 0.69}
\definecolor{seabornorange}{rgb}{0.867, 0.518, 0.322}
\definecolor{seaborngreen}{rgb}{0.333, 0.659, 0.408}

\newcommand{\X}{\mathrm{X}}
\newcommand{\B}{\mathrm{B}}
\newcommand{\A}{\mathrm{A}}
\newcommand{\Y}{\mathrm{Y}}
\newcommand{\E}{\mathrm{E}}
\newcommand{\F}{\mathrm{F}}
\newcommand{\D}{\mathrm{D}}
\newcommand{\Z}{\mathrm{Z}}

\journal{Elsevier}

\begin{document}

\begin{frontmatter}






\title{A Blind Source Separation Framework to Monitor Sectoral Power Demand from Grid-scale Load Measurements}


\affiliation[mox]{organization={MOX, Department of Mathematics, Politecnico di Milano},
            addressline={Piazza Leonardo da Vinci 32}, 
            city={Milano},
            postcode={20133}, 
            state={MI},
            country={Italy}}

\affiliation[deng]{organization={Department of Energy, Politecnico di Milano},
            addressline={Via Lambruschini 4a}, 
            city={Milano},
            postcode={20156}, 
            state={MI},
            country={Italy}}

\affiliation[terna]{
            organization={Statistics Group, Terna S.p.A},
            addressline={Viale Egidio Galbani, 70}, 
            city={Rome},
            postcode={00156}, 
            state={RM},
            country={Italy}}

\author[mox]{Guillaume Koechlin\corref{cor1}}
\cortext[cor1]{Corresponding author}
\ead{guillaume.koechlin@polimi.it}

\author[deng]{Filippo Bovera}

\author[terna]{Elena Degli Innocenti}

\author[terna]{Barbara Santini}

\author[terna]{Alessandro Venturi}

\author[terna]{Simona Vazio}

\author[mox]{Piercesare Secchi}

\begin{abstract}
\rev{As demand-side flexibility becomes increasingly necessary to integrate variable renewable energy, understanding electricity demand composition across different grid levels is essential. However, at regional and national scales, visibility into the relative contributions of different consumer categories remains limited due to the complexity and cost of collecting end-use consumption data.
To address this challenge, we propose a blind source separation framework to disaggregate open-access high-voltage grid load measurements into sectoral contributions. The approach relies on a constrained variant of non-negative matrix factorization, termed linearly-constrained non-negative matrix factorization (LCNMF), which allows prior information to be incorporated as linear constraints on the factor matrices, thereby providing weak supervision of the separation process. 
The framework is evaluated using Italian national load data from 2021 to 2023. Results demonstrate the identifiability of residential, services, and industrial load components and provide monthly sectoral consumption estimates consistent with reported statistics. The proposed method is generalizable and applicable to load disaggregation problems across multiple grid scales where disaggregated measurements are unavailable.}
\end{abstract}




\end{frontmatter}



\section{Introduction}
\label{intro}

\subsection{Motivations}
Monitoring and understanding electricity consumption dynamics is essential for power system operation. At the national level, electricity consumption (and generation) is typically monitored by the transmission system operator (TSO) both at the level of geographical zones, such as states, regions or power zones, and across different sectors often grouped in three or four categories: industry, commercial \& public services (or simply \textit{commercial} or \textit{services}), residential (or \textit{households}) and sometimes agriculture \cite{iea_key_2021, mor_datadriven_2021}.
While zone-level consumption is directly accessible to the TSO at high temporal resolution via measurements on the high-voltage network, sector-level consumption is considerably more difficult to monitor.
It must be retrieved from the medium- and low-voltage networks,\footnote{With the exception of a few industrial consumers directly connected to the high-voltage grid.} which fall outside the TSO's scope and are managed by numerous distribution system operators (DSOs). This process thus involves aggregating numerous building-level load measurements, which are not centrally collected. As a result, sector-specific consumption data are often available only at very poor temporal resolution, mostly on an annual basis, and with significant reporting delays \cite{voulis_understanding_2018}.

However, estimating sector-specific consumption at finer temporal resolutions, such as monthly or daily, or even directly monitoring the load at the sector level, is of particular interest to TSOs and other power system actors for two reasons: 

First, it obviously provides precious knowledge on sectors themselves.
For instance, global sectors' typical load profiles and granular consumption can be used for the calibration and validation of building stock energy models, especially in top-down approaches \cite{langevin_developing_2020}. It is also valuable for demand-side management as it allows to understand load composition during peak hours -- which may vary depending on the type of day or the season -- and therefore help designing efficient sector-specific demand response programs \cite{wang_load_2015}. Real-time consumption monitoring can also help rapidly detect and quantify how the different sectors respond to economic shocks (e.g., energy price spikes) \cite{mollerandersen_changes_2024, thoma_electrical_2004}, policy interventions (e.g., lockdowns) \cite{wang_electricityconsumption_2021, krarti_review_2021, vanzoest_evaluating_2023} or extreme weather events (e.g., heat waves) \cite{zhang_extreme_2022}, which is increasingly important in a context of rising demand-side and generation flexibility needs. Past and current efforts on sectors' power consumption monitoring rely on small consumer samples, not always representative nor accessible \cite{voulis_understanding_2018, gerossier_novel_2017}, and only available a posteriori. Finally, sector-level demand at higher frequency is obviously not only useful to the power sector but also to economic institutions and policymakers to monitor the national economic activity \cite{kraft_relationship_1978} with more precision.

Second, it can enhance aggregate load modeling and forecasting \cite{andersen_long_2013, andersen_longterm_2013, diaz-iglesias_shortterm_2025}. A significant portion of the variability in the national load curve arises from the dynamic contributions of different sectors \cite{andersen_longterm_2013}. For example, weekday load profiles are typically flatter than those on weekends due to a stronger industry contribution, whose consumption pattern is rather stable over the day. It may also improve the modeling of weather or calendar effects on the aggregate load curve, as different load components (i.e. sectors) may respond differently to specific conditions. For instance, the households and services sectors may show higher temperature sensitivity in summer due to widespread air conditioning usage, compared to the industrial sector \cite{pagliarini_outdoor_2019}. Additionally, the long-term evolution of sector-specific load profiles may be easier to model and project in scenarios -- especially in the context of electrification -- than the aggregate national load \cite{andersen_long_2013, andersen_differentiated_2014, mollerandersen_changes_2024}.

\subsection{Objectives of the Work}

In this work, we demonstrate that, despite the absence of actual measurements, it is possible to monitor the hourly consumption of the residential, commercial and industrial sectors in real-time from the aggregate open-access TSO load measurements, by applying signal separation techniques. \rev{The signal separation framework we propose enables both the extraction of sector-specific daily load profiles and the nowcasting of daily or monthly sectoral consumption}. We formulate this unsupervised load disaggregation task as a \textit{blind} source separation (BSS) problem, showing that the incorporation of sector-level information usually available -- such as annual demand and monthly consumption proxies -- as boundary conditions allows to identify characteristic sectors' load profiles and obtain an accurate sector-wise disaggregation. The BSS problem is formalized as a constrained variant of non-negative matrix factorization (NMF) \cite{lee_learning_1999}, which is termed linearly-constrained non-negative matrix factorization (LCNMF), and solved using a modified version of the multiplicative updates algorithm \cite{lee_algorithms_2000}. We apply this method to Italian national load data from 2021-2022 and show that the identified sector profiles for this period allow to accurately disaggregate in real-time new load curves for which no sector-level information is available, using 2023 data for validation.

\subsection{Paper structure}

The remainder of this paper is structured as follows. Section \ref{sec:lit} provides a detailed literature review and explains our contributions. Section \ref{sec:meth} details the methodology, beginning with a description of the data sources (\ref{sec:data}) and a qualitative introduction to the problem (\ref{subsec:quali}) before presenting the mathematical and computational details (\ref{subsec:model} and \ref{subsec:sol}). Section \ref{sec:res} presents the results, and Section \ref{sec:conc} concludes with a discussion and final remarks.

\section{Literature review}
\label{sec:lit}

Statistical modeling and analysis of load curves is a major topic in power systems. A literature overview suggests that most works belong to one of the three following categories:
\begin{enumerate}
    \item \textbf{Load forecasting}: The problem is to predict future load at various network scales from past measurements and influencing factors such as weather and calendar effects. Though the results of the present work may be relevant for such purposes, the problem we solve is not a forecasting one and the corresponding literature is therefore not further presented here.
    \item \textbf{Load profiling/classification}: This category regards the analysis of typical patterns and similarities between load curves observed in different locations, for different day types, seasons or categories of consumers. The comparison with such works is relevant since our method involves identifying sector-specific load profiles.
    \item \textbf{Load disaggregation}: We refer to load disaggregation as the task of separating unobserved devices, end-users or consumer groups contributions to observed aggregated load measurements.  This is precisely the problem of interest. Since nearly all load disaggregation works are limited to device-level decomposition of buildings consumption, we selected those whose methodological framework is close to ours and this part of the review is essentially on the technical side.
\end{enumerate}

\rev{In addition, we demonstrate that this work brings a genuine methodological novelty by extending the discussion to all contributions that have proposed constrained non-negative matrix factorization methods, regardless of the application.}

\subsection{Load profiling and classification}

\smallskip

In load profiling and classification, a distinction can be made according to whether load measurements or models corresponds to end-users or an aggregated level (such as district or regional/national level).


At the end-user level, typical residential, commercial and industrial load profiles can be inferred by averaging individual consumers' load curves within each sector. This is typically done with a manual weekday/weekend, season, region and household characteristics stratification or in a more data-driven fashion with clustering techniques. As highlighted in \cite{voulis_understanding_2018}, thanks to the large diffusion of smart meters, the residential sector has received most attention in the literature \cite{paatero_model_2006, rasanen_databased_2010, teeraratkul_shapebased_2018, andersen_residential_2021, michalakopoulos_machine_2024, jin_characterizing_2024}, while fewer works studied commercial and industrial consumers \cite{jardini_daily_2000, bellinguer_elmas_2023}. Our study, instead, considers aggregate country-level profiles and their temporal variations rather than within-sector heterogeneity.

Regarding the aggregate level, most works use a so-called bottom-up approach \cite{langevin_developing_2020}: The load profile of the aggregation level is built by aggregating load profiles of different consumer categories with weights reflecting the local consumer mix. For instance, \cite{voulis_understanding_2018} builds neighborhood, district and municipality-level load profiles, which are shown to be well-described by a three-level ``residential", ``business", and ``mixed" classification. Their results show that at aggregate levels, only global sector categorization really makes the difference, rather than finer ones. Their study however lacks the inclusion of the industrial component, which is instead present in our work. \cite{andersen_longterm_2013} aggregates load curves of 4500 end-users in Denmark to derive national profiles for residential, commercial, public services, industry and agriculture sectors, finding strong weekly and seasonal variation except for industry. Later work \cite{mollerandersen_changes_2024} shows that sector profiles remain relatively stable over the 2019-2022 period. Finally, \cite{andersen_long_2013, andersen_differentiated_2014} adopt more of a top-down approach as they estimate the consumer mix in local areas in Denmark by combining an end-users load profiles dataset and transformer station aggregate measurements. Note that, similarly to \cite{voulis_understanding_2018}, they rely on the assumption that load profiles of specific consumer categories are constant across locations and, differently from us, do not properly model sectors' load profiles variations over time. Moving to the highest aggregation level, \cite{behm_how_2020} proposed a methodology to build weather-dependent load profiles for European countries. They however do not address the weather-independent temporal variations of load profiles and sector-specific consumption patterns.

\smallskip

\subsection{Load disaggregation}


Prior research on load profiling and classification has consistently shown that aggregate curves from the substation to the national level contain identifiable sector components. However, while building-scale load disaggregation is a flourishing field of study, which we will discuss first, disaggregation at higher grid scales remains largely unexplored. We will review the few existing works on this topic in a subsequent section.

\subsubsection{Non-intrusive load monitoring}

First introduced in \cite{hart_nonintrusive_1992}, building-scale load disaggregation is referred to as non-intrusive load monitoring (NILM) or sometimes energy disaggregation. The purpose of NILM is to monitor a building's load at the appliance level without the necessity of installing dedicated sensors on every single device -- which in this case is referred to as intrusive load monitoring (ILM) -- but with a unique building-level energy meter, reducing therefore costs and addressing privacy issues.
While this work is not about NILM, it's actually solving a similar load disaggregation problem at a higher level of the network, with the difference that it separates consumer categories rather than individual appliances. Because of this conceptual similarity, we believe it's relevant to compare our methodological approach to established NILM techniques.

\smallskip

According to a recent review by \cite{schirmer_nonintrusive_2023}, NILM methods fall into three primary families: machine learning, pattern matching, and source separation. It's important to note that most of these approaches, especially those in the first two families, are supervised, meaning they require access to ground truth, appliance-level measurements to which a learning procedure is applied. In contrast, our method operates in an unsupervised fashion, directly estimating sector load profiles from a composite signal without needing any ground truth measurements.

In the source separation methods, three different categories are distinguished: independent component analysis (ICA) \cite{zhu_load_2014, semwal_practicability_2013, herath_assessment_2023}, non-negative matrix or tensor factorization (NMF/NTF) \cite{figueiredo_electrical_2014, matsumoto_energy_2016, rahimpour_nonintrusive_2017, miyasawa_energy_2019, zarabie_l0norm_2019} and sparse component analysis (SCA) \cite{singh_analysis_2019, piga_sparse_2016, kolter_energy_2010}. All three methods have in common that they solve a low-rank matrix (or tensor) approximation problem, differing however in the assumptions they rely upon and the factor matrices properties they look for. A key point is that they are in practice unsuccessful when they are used in a fully blind, i.e. unsupervised, setting, as in this case there is often non-uniqueness of the solution and high dependence on the initialization strategy. In fact, \cite{schirmer_nonintrusive_2023} states that source separation methods developed for NILM, despite being originally thought as unsupervised, are actually ``semi-unsupervised" learning procedures, as they all require some degree of prior knowledge which can range from ground truth source signal data to little contextual or structural information on the sources or the mixture.

Notably, both \cite{kolter_energy_2010} (SCA-based) and \cite{rahimpour_nonintrusive_2017} (NMF-based) rely on training data consisting of individual appliance-level measurements. Their methods involve modeling each source load (appliance) using basis functions -- predefined in \cite{rahimpour_nonintrusive_2017} and learned in \cite{kolter_energy_2010} -- which are then used to disaggregate the aggregate load. Interestingly, \cite{rahimpour_nonintrusive_2017} formulates an NMF problem with a linear constraint on the weights matrix, similar in form to ours. However, they treat this as a non-negative least squares (NNLS) problem, updating only the weights while keeping the basis fixed.

Finally, \cite{wytock_contextually_2014} propose a contextually-supervised source separation framework. They incorporate prior information about appliances load via penalty terms, allowing for weak supervision in the absence of full training data. However, their paper does not provide details on how the resulting penalized matrix factorization problem is solved algorithmically.

\subsubsection{Beyond the building level}

To the best of our knowledge, only two works tackled a load disaggregation problem beyond the building scale, among which only one concerning a disaggregation in consumer categories:
\begin{itemize}

\item A recent study\footnote{In which the authors confirm the literature gap we identify.} \cite{dampeyrou_unsupervised_2024} proposing an unsupervised method for separating the weather- and calendar-dependent components of the national load curve. Similarly to us, they don't rely on the observation of disaggregated measurements for model training. Nevertheless, their methodological framework does not resembles ours as they estimate the separate components as non-linear functions of weather and calendar variables with a deep-learning method. The authors themselves note a key limitation: the resulting decomposition lacks meaningfulness and practical utility. In contrast, our approach's decomposition is both identifiable and meaningful, as prior research on load profiling showed.

\item Another, more relevant work is \cite{gerossier_novel_2017}, proposing a source separation framework to estimate consumer category profiles from transformer substation load data. Their method uses a customer database containing annual consumption for each consumer category, which is available for every area covered by the substations. The main difference on the application side regards the fact that they work at the low-medium voltage vs high-voltage network scale in our case. On the methodological side, since they already know the consumer categories weights in the aggregate curves, they solve a non-blind source separation problem -- which boils down to NNLS -- for which a global minimum exists. Our source separation problem instead is truly blind, as consumer categories weights are unknown and must be estimated simultaneously with the source profiles. This optimization problem is more complex as it is non-convex and therefore no global minimum can be found \cite{gillis_nonnegative_2020}. Though valuable for the specific problem tackled by the authors, their study has several limitations affecting results validity and its generalizability to similar yet different load disaggregation problems like ours:

\begin{enumerate}
    \item It relies on a non-typical customer database -- belonging to the French DSO -- providing very local consumer information which may not be available for other applications. Our method, conversely, uses widely available country-level annual and monthly consumption data.
    \item As noted by the authors in the conclusion, they do not have an automatic way to determine the number of source profiles that can be extracted from the aggregate curves. In our case, though we look for a three-level decomposition, we model the aggregate load with a higher number of source profiles, directly inferred from the data.
    \item By assigning a single and fixed profile to each consumer category, they assume that average consumption behaviors at this geographical scale are temporally and spatially invariant. However, \cite{andersen_differentiated_2014} highlights that, while this is a reasonable assumption for households, the industry profile is very likely to differ between areas depending on the local industrial mix. Moreover, the results of \cite{andersen_longterm_2013} clearly show that households and services profiles change over the year. The framework of the present work, instead, enables flexible profiles that can account for intra-week and seasonal variations.
    \item They only test the ability of their model to accurately represent a new transformer aggregate load curve but they don't validate the accuracy of their decomposition. This is obviously due to the lack of ground truth measurements, like us, but in our case we validate the disaggregated estimates at the month level against sector-level consumption indicators and compare the load profiles with those reported in the literature.
\end{enumerate}

\end{itemize}

\subsection{\rev{Constrained NMF}}

\rev{Beyond contributions specifically addressing load disaggregation problems, several \textit{constrained} NMF (CNMF) methods have been proposed in the literature, most of which extend the original multiplicative updates (MU) algorithm introduced in \cite{lee_algorithms_2000}. A recent review can be found in \cite{guo_rise_2024}. However, most papers tackle highly specific cases which are not generalizable to our problem. For instance, \cite{liu_semisupervised_2006, liu_constrained_2012} introduce a so-called CNMF method which is in reality more akin to nonnegative matrix \textit{tri}-factorization (NMTF).}

\rev{Several contributions -- mainly in the hyperspectral unmixing literature -- focus on \textit{sparse} or \textit{smooth} NMF (which can be both considered as belonging to CNMF, c.f. \cite{guo_rise_2024}), where a $L^p$-norm or a more elaborate regularization term is incorporated into the objective function \cite{pauca_nonnegative_2006, jia_constrained_2009, liu_approach_2011, chen_nonnegative_} to promote sparsity or smoothness in the factor matrices. In a different direction, \cite{hosseinzadehaghdam_novel_2022} proposes a specific CNMF method -- classified as NMF \textit{in manifolds} in \cite{guo_rise_2024} -- aiming at preserving pairwise similarities between observations in the reduced space. Finally, \cite{peng_multiplicative_2012} provide theoretical grounding for our LCNMF method by showing that the MU algorithm can be trivially tweaked to solve NMF with additional linear constraints on the factor matrices, although the modified MU are not explicitly detailed in their paper.}

\rev{Consequently, this work introduces a novel algorithmic framework for solving a general NMF problem with linear constraints on the factor matrices.}


\subsection{Research gap and main contributions}

All aforementioned studies on aggregate load modeling and profiling rely on the availability of large, costly, end-users load samples or atypical consumer information datasets, making them poorly generalizable. Besides, most of them put efforts in the definition of granular load-based consumer classifications while failing to account for the temporal variations of daily profiles. Moreover, none of them address the problem of monitoring the total consumption of an entire consumer group, such as households, services or industry at the national level and at high frequency. Despite studies showing that highly aggregate load curves variations reflect a dynamic composition of distinctive sectors consumption profiles, no work explored the potential of load disaggregation methods for large-scale sectors consumption monitoring. \rev{Finally, no existing methodological framework makes it possible to solve a general NMF problem with linear constraints, such as the one considered in this paper.} To address these limitations, this paper makes the following contributions:

\begin{enumerate}[(i)]
    \item \rev{We are the first to propose a transmission-level load disaggregation method for monitoring sectoral power demand from national high-voltage grid load measurements.}
    \item We introduce a novel top-down approach for identifying country-level sectors daily load profiles and their temporal variations which does not require the availability of consumer-level samples and relies on open-access TSO data only.
    \item \rev{We provide an algorithmic framework to solve a general NMF problem with linear constraints on the factor matrices, which can be applied to unsupervised load disaggregation problems at any scale of the power grid.}
\end{enumerate}



\section{Methodology}
\label{sec:meth}

\begin{figure}[h!]
\centering
\resizebox{0.9\textwidth}{!}{\usetikzlibrary{positioning, arrows.meta}


\begin{tikzpicture}[
  principal/.style={rectangle, draw, fill=blue!10, thick, text centered, font=\large},
  data/.style={rectangle, draw, fill=blue!5, text centered},
  algo/.style={circle, draw, fill=gray!10, thick, text centered, minimum size=2cm, font=\small\bfseries},
  output/.style={rectangle, draw, fill=red!10, text centered, font=\large},
  arrow/.style={-Stealth, thick}
]

\node[principal, align=center] (load) {
\textbf{Total Load} \\
\textit{Resolution}: Hourly \\
\textit{Granularity}: Aggregated};

\node[algo, align=center, below=0.8cm of load] (algorithm) {
Linearly-Constrained \\ Nonnegative \\ Matrix Factorization \\ (LCNMF)};

\node[data, align=center, left=0.8cm of algorithm] (consumption) {
\textbf{Annual Sector Demand} \\ \textbf{(ASD)} \\
\textit{Resolution}: Yearly \\
\textit{Granularity}: Sector-level};

\node[data, align=center, right=0.8cm of algorithm] (indicators) {
\textbf{Monthly Sector Indicators} \\ \textbf{(MSI)} \\
\textit{Resolution}: Monthly \\
\textit{Granularity}: Sector-level};

\node[output, align=center, below=0.8cm of algorithm] (result) {
\textbf{Disaggregated Load} \\
\textit{Resolution}: Hourly \\
\textit{Granularity}: Sector-level};

\draw[arrow] (load) -- (algorithm);
\draw[arrow] (consumption) -- (algorithm);
\draw[arrow] (indicators) -- (algorithm);
\draw[arrow] (algorithm) -- (result);

\end{tikzpicture}

}
\caption{Diagram showing the load decomposition process. The input data are represented in blue rectangular boxes (lighter blue for secondary inputs) while the output of the LCNMF disaggregation procedure is represented as a red rectangular box.}
\label{fig:data-wf}
\end{figure}
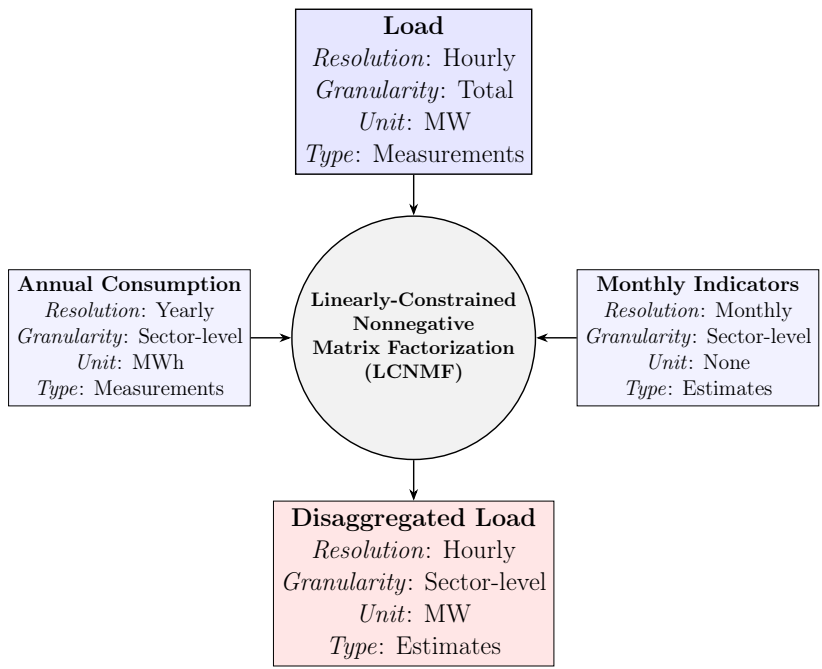

\noindent Fig. \ref{fig:data-wf} provides an input-output diagram of the load decomposition process.

\subsection{Data}
\label{sec:data}
The consumption data we aim to decompose come from the Italian TSO's (Terna) hourly measurements of the national load in Italy between 2021 and 2023
\cite{terna_total_2024}.
Additionally, available information on sectors' electricity consumption is used.
Specifically, we use two sector-level data sources:
\begin{enumerate}
\item[(i)] The yearly consumption per sector from Terna's annual consumption report \cite{terna_consumi_2024}, which we refer to as the \textit{annual sector demand} (ASD). 
\item[(ii)] Three unitless indicators \cite{terna_enel_2024, terna_imcei_2024, terna_imser_2024} \rev{used by Terna}, describing the monthly evolution of the household, industry\footnote{Which also includes the agricultural sector. Given that the agricultural sector's weight in total consumption is negligible compared to the three other sectors, and since we have no suitable indicator to act as a proxy for its monthly consumption, we have chosen to merge this sector with the industrial one (as we believe their consumption patterns are most similar)} and services sectors consumption, designated as \textit{monthly sectoral indicators} (MSI). 
\end{enumerate}

\noindent Table \ref{tab:data_sources} shows a tabular view of the three different data sources and preprocessing steps applied.

\begin{table}[htbp]
\centering
\small
\rev{
\begin{tabular}{lllll}
\toprule
\textbf{Data} & \textbf{Resolution} & \textbf{Unit} & \textbf{Source} & \textbf{Preprocessing} \\ \midrule
Load & Hourly & MW & \cite{terna_total_2024} & \makecell[l]{Daylight saving time (DST) \\ adjustment\tablefootnote{\rev{The second occurrence of the duplicate timestamp (02:00) on hour gain days is deleted and the missing timestamp (02:00) on hour loss days is added with the load value observed at 01:00.}}} \\ 
\addlinespace
ASD & Yearly & MWh & \cite{terna_consumi_2024} & \makecell[l]{Merge of Agriculture into Industry \\ and rescaling to match load data\tablefootnote{\rev{Note that the decomposition, originating from load data, will inherently overestimate the sectors' consumption as it disaggregates a quantity that includes network losses, being computed from the generation side of the grid. To address this issue, one can either subtract the losses to the load data or add the losses proportionally to the ASD, before running the separation process.}}} \\ 
\addlinespace
MSI & Monthly & None & \cite{terna_enel_2024, terna_imcei_2024, terna_imser_2024} & None \\ \bottomrule
\end{tabular}
}
\caption{Data sources and preprocessing steps}
\label{tab:data_sources}
\end{table}

\subsection{Qualitative description}
\label{subsec:quali}

Blind source separation (BSS) is a technique used to separate a set of mixed signals into their original, independent source components, without prior knowledge of the mixing process or the source signals.
In our load disaggregation problem, we observe each day a load curve having a specific shape resulting from the mixture of the sectors' load profiles -- which can be seen as the source signals. We aim to leverage the fact that some sectors are predominant on certain days to identify their load profile, estimate their contribution to the overall mixture, and reconstruct their load signal. However, beyond this necessity to have mixture weights, i.e. sector importance, that show some degree of variability, BSS assumes that sectors have distinct and time-invariant load profiles. While it is reasonable to expect that the typical load curve of one sector differs from another, it is clear that these profiles can vary based on factors such as the day of the week (weekday vs. weekend) or the season (winter vs. summer) as shown in \cite{andersen_longterm_2013}. Therefore we cannot consider the sectors' load profiles as fixed and constant over time. To address this limitation, we assume that each sector’s load profile is itself a mixture of multiple underlying sources, which may or may not be directly interpretable, and that these sources meet the conditions for separation. Note that the number of sources present in the aggregated signal can be inferred from the data with dimensionality reduction techniques such as principal component analysis (PCA). By identifying the contribution of these sources to the overall load signal, we can estimate the contribution of each sector by aggregating the corresponding sources.

\begin{figure}[h]
    \centering
    \includegraphics[width=\linewidth]{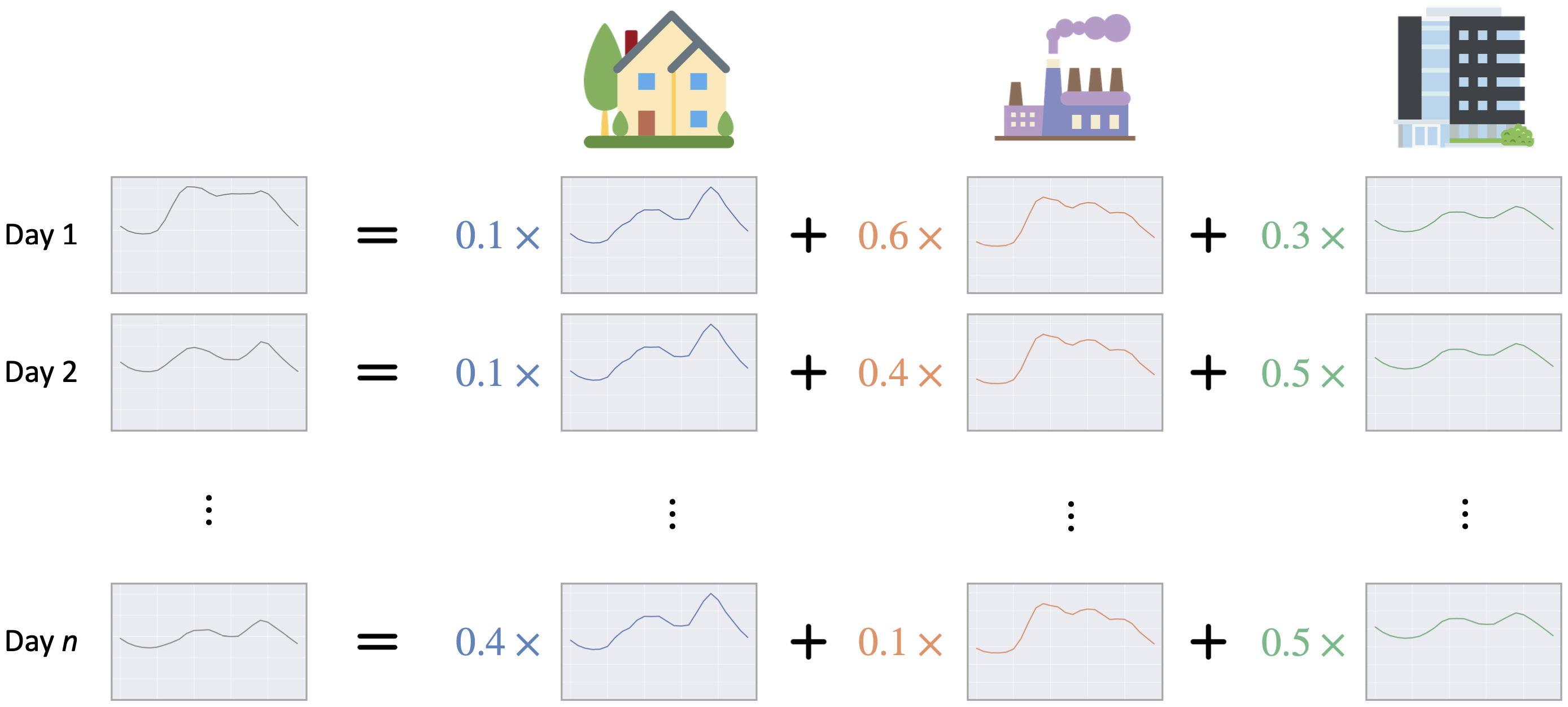}
    \caption{Illustration of the load disaggregation problem. In this example, each sector is associated to one source only.}
    \label{fig:bss_load}
\end{figure}

However, without incorporating prior information on the sectors profiles or weights, the likelihood of finding the desired decomposition -- where the sources are precisely the “voices” of our sectors -- is quite low. Fortunately, we do have valuable data that can guide the separation process: the ASD and MSI. The ASD provides information on sectors' total annual consumption, while the MSI indicates how this consumption is distributed over the 12 months. Hence an estimate of each sector’s monthly consumption can be computed, and this can be compared to the results of the decomposition, aggregated at the month level. By constraining the decomposition to produce monthly consumption estimates that align with these known values, we can identify sources that are hierarchically linked to the sectors and hence reconstruct the sectors' load.

Once we have identified the sources, any \textit{new} set of daily load curves can be decomposed into contributions from these sources without requiring any further information about sectors' consumption (which, by contrast, is necessary for the initial source identification). In other words, we can decompose new load data using only the load curves themselves (and the identified sources, obviously). The source separation process becomes “non-blind” because we only need to estimate the weights of the observed mixtures. To validate this nowcasting predictive approach, we decompose the load curves for the year 2023 using the sources estimated on 2021-2022, without using any further sector-level data. We then compute the predicted monthly consumption and compare it to the MSI for verification.

The next two subsections dive into the mathematical details of what was just described. 

\subsection{Model}
\label{subsec:model}


Let $L_i: [0,T] \rightarrow \mathbb{R}^+$ represent the system's\footnote{The national power grid in our case} load as a function of time during a day $i \in \{1, \dots, n\}$. The total energy consumption $e_i$ during day $i$ is given by:
$$e_i = \int_{0}^{T} L_i(t)dt = \lVert L_i \rVert_{1}$$
We define $X_i: [0,T] \rightarrow \mathbb{R}^+$ as the load shape curve observed on day $i$:
$$L_i = e_i \cdot X_{i}$$
In other words, $X_i$ is the normalised load curve of day $i$ such that it corresponds to a total consumption of 1. For notation convenience, we may assimilate $X_i$ to an element of $\mathbb{R}_+^p$, corresponding to evaluations of the underlying function at $p$ regularly-spaced time instants spanning the interval $[0,T]$. For example, $p=25$ corresponds to evaluations at 00:00, 01:00, \dots, 24:00.


\medskip

Each curve $X_i$ is assumed to be a mixture of $K$ \textit{sources} $S_1, \dots, S_K$ with $K \ll n$:
\begin{equation}
    \label{sys}
    \begin{cases}
    X_{1} &= c_{11}S_{1} + c_{12}S_{2} + \dots + c_{1K}S_{K} + \epsilon_1 \\
    X_{2} &= c_{21}S_{1} + c_{22}S_{2} + \dots + c_{2K}S_{K} + \epsilon_2 \\
    &\vdots \\
    X_{n} &= c_{n1}S_{1} + c_{n2}S_{2} + \dots + c_{nK}S_{K} + \epsilon_n
\end{cases}
\end{equation}
where for \(i \in \{1,\dots, n\}\) and \(k \in \{1,\dots,K\}\), the coefficient $c_{ik}$ represents the \emph{concentration} of source $k$ in the load (shape) curve $X_i$  and $\epsilon_1, \dots, \epsilon_n$ are additional noise terms. The profiles $S_1, \dots, S_K$ and the concentrations \(c_{ik}\)'s are constrained to be positive. Additionally, we impose that $S_1, \dots, S_K$ integrate to 1 (as the $X_i$'s) and thus
$c_{i1} + \dots + c_{iK} = 1$, for all $i \in \{1,\dots, n\}$.
The system of equations \eqref{sys} can be re-written in a matrix form:
\begin{equation}
    \X \approx CS
\label{nmf}
\end{equation}
where $\X$ is the $n \times p$ load curves matrix, $C$ the $n \times K$ concentrations matrix and $S$ the $K \times p$ sources matrix:

$$
\X = \begin{bmatrix} X_1 \\ X_2 \\ \vdots \\ X_n \end{bmatrix} \! , \;
C = \begin{bmatrix}
    c_{11} &  c_{12}  & \cdots &  c_{1K}
\\   c_{21} &  c_{22}  & \cdots &  c_{2K}
\\  \vdots & \vdots & & \vdots
\\    c_{n1} &  c_{n2}  & \cdots &  c_{nK}
\end{bmatrix} \; \text{and} \ S = \begin{bmatrix} S_1 \\ S_2 \\ \vdots \\ S_K \end{bmatrix}.$$

\medskip

The load can hence be decomposed by finding a pair of matrices $\hat{C}$ and $\hat{S}$ whose product $\hat{C}\hat{S}$ best approximates the data $\X$. The estimated consumption $\hat{e}_{ik}$ of source $k$ during day $i$ is then deduced by multiplying the estimated concentration $\hat{c}_{ik}$ with the total consumption $e_i$:
$$\hat{e}_{ik} = \hat{c}_{ik} e_i$$

$K$ can be chosen by estimating the underlying dimension $d$ of the space in which the load curves $\X$ are embedded, e.g. with functional principal component analysis (FPCA) \cite{ramsay_functional_2005}. Given that the fitted curves $CS$ are convex linear combinations of the sources and live therefore in a $(K-1)$-dimensional linear subspace of $\mathbb{R}_+^p$, we must choose $K=d+1$.

\smallskip

The low-rank matrix approximation problem of equation \eqref{nmf}, when both matrices are constrained to have all their coefficients positive, is known as \textit{Non-negative Matrix Factorization} (NMF), with the difference that we have an additional constraint due to the fact that the profiles contained in $S$ must
integrate to 1. Moreover, as stated in the previous section, we would like to add a constraint on the solution such that it is consistent with the ASD-MSI sectors' monthly consumption estimates.
To formalise this, let $E = [e_1 \; e_2 \, \dots \, e_n]^\prime $ be the $n$-dimensional vector of daily consumption, for which we consider a block-form notation in which the vector elements are grouped by month:
$$E = \begin{bmatrix} E_1 \\ E_2 \\ \vdots \\ E_m \end{bmatrix}$$
where $m$ is the number of months in the period considered.
We additionally define a $m \times n$ matrix $\B$ and re-write the $n \times K$ matrix $C$ of concentrations in a similar block-form manner:

$$
\B = \begin{bmatrix}
    E_1^\prime &  0  & \cdots &  0
\\   0  & E_2^\prime & \cdots &  0
\\  \vdots & \vdots & \ddots & \vdots
\\   0  &  0  & \cdots & E_m^\prime
\end{bmatrix}
 \; \text{and}
\ C = \begin{bmatrix} C_1 \\ C_2 \\ \vdots \\ C_m \end{bmatrix}$$
\noindent Multiplying the two matrices gives us the $m \times K$ matrix of the monthly consumption of the $K$ sources:
$$\B C = \begin{bmatrix} E_1^\prime C_1 \\ E_2^\prime C_2 \\ \vdots \\ E_m^\prime C_m \end{bmatrix}$$

\medskip

\noindent The problem now concerns mapping the $K$ sources to the $g$ sectors. We may assign more sources to sectors where we expect the characteristic load profile to vary more significantly throughout the year.
We formalise this mapping by defining a surjection $\sigma: \{1,\dots,K\} \rightarrow \{1,\dots,g\}$ and the associated $K \times g$ matrix $\A$ such that, for source $i \in \{1,\dots,K\}$ and sector $j \in \{1,\dots,g\}$:
$$\A_{i,j} = \begin{cases}
      1 & \text{if}\ j=\sigma(i) \quad \text{i.e. if $i$ is a source of sector $j$}\\
      0 & \text{otherwise}
    \end{cases}$$
For instance, in the case $K=5$ and $g=3$, attributing source 1 to sector 1, sources 2 and 3 to sector 2, and sources 4 and 5 to sector 3, we have:

$$\A = \begin{bmatrix}
    1 & 0 & 0
 \\ 0 & 1 & 0
 \\ 0 & 1 & 0
 \\ 0 & 0 & 1
 \\ 0 & 0 & 1
\end{bmatrix}$$
We can finally obtain the $m \times g$ matrix $\Z$ of the estimated monthly consumption per sector:
$$\Z = \B C\A$$
On the other side, let $\Y=[Y_1 \dots Y_g]$ be the $m \times g$ matrix whose columns \(Y_j\)'s
represent the MSI, which, as highlighted in subsection \ref{subsec:quali}, only carry information on the \textit{shape} that the columns of $\Z$, seen as monthly time series, must assume. Intuitively, we would like to drive the decomposition to yield a $\Z$ matrix with  columns as \textit{correlated} as possible with those of $\Y$, regardless of the difference in scale. However, the introduction of a penalty term based on the Pearson correlation coefficient -- or other relevant similarity measure quantifying statistical dependence -- between $\Z_j$  and $\Y_j$ in the loss function may be computationally intractable.
To address this issue, our approach is to use the ASD to \textit{rescale} the matrix $\Y$ so that it can directly be compared to $\Z$.
More specifically, let $w_j$, for $j \in \{1, \dots, g\}$, be the total consumption given by the ASD
of sector $j$ over the period considered (which must necessarily be composed of entire years). Note that we should have $w_1 + \dots + w_g=\lVert E \rVert_{1}$.
Performing the following scaling operation, for all $j \in \{1, \dots, g\}$:
$$Y_j \leftarrow \frac{Y_j}{\lVert Y_j \rVert_{1}}w_i$$
$\Y$ is now exactly an estimate of what $\Z$ should be.
Note that the boundary condition $Y_1 + \dots + Y_g = [\lVert E_1 \rVert_{1}, \lVert E_2 \rVert_{1}, \dots, \lVert E_m \rVert_{1}]^\prime $, i.e. summing the sectors' consumption should give the total consumption at the month level, may not be fully satisfied but we can similarly rescale $\Y$ row-wise to correct this effect. 
\bigskip

\noindent We can write down the final optimisation problem:
\begin{equation}
\label{obj}
\begin{aligned}
    (\hat{C},\hat{S})=\argmin_{C \in \mathbb{R}_+^{n \times K},\, S \in \mathbb{R}_+^{K \times p}} & \, \text{dist}(\X, \, C S)^2 \\
    \text{s.t.} \qquad &  \hspace{-1.5em}\; \text{B} C \text{A} = \text{Y} \\
     &  \hspace{-1.5em}\; S \, \mathbf{1}_p = \mathbf{1}_K\\
\end{aligned}
\end{equation}
where $\text{dist}()$ is some distance measure on $\mathbb{R}_+^{n \times p}$ and $\mathbf{1}_r$ is the $r$-dimensional vector of ones.

\subsection{Solution method}
\label{subsec:sol}
The Multiplicative Updates (MU) algorithm introduced by Lee \& Seung \cite{lee_algorithms_2000} allows to find a local minimum to the ordinary NMF problem, formulated with a distance measure being the metric induced by the Frobenius norm (or another distance measure similar to the Kullback-Leibler divergence). Essentially, from arbitrary initial values, the matrices $C$ and $S$ get alternatively updated by multiplying their current value by a factor that depends on the quality of the approximation, until convergence.

\smallskip

We propose a modified version of the MU algorithm to solve the \textit{linearly constrained} non-negative matrix factorization (LCNMF) problem, i.e., NMF with additional equality constraints that take a linear form in the factor matrices, generalizing \eqref{obj}.
Indeed, given any $\B \in \mathbb{R}_+^{m \times n}$, $\A \in \mathbb{R}_+^{K \times g}$, $\Y \in \mathbb{R}_+^{m \times g}$, $\F \in \mathbb{R}_+^{l \times K}$, $\D \in \mathbb{R}_+^{p \times q}$, $\Z \in \mathbb{R}_+^{l \times q},$ let
\begin{equation}
\label{lcnmf}
\begin{aligned}
    (\hat{C},\hat{S})=\argmin_{C \in \mathbb{R}_+^{n \times K},\, S \in \mathbb{R}_+^{K \times p}} & \, \text{dist}(\X, \, C S)^2 \\
    \text{s.t.} \qquad &  \hspace{-1.5em}\; \text{B} C \text{A} = \text{Y} \\
     &  \hspace{-1.5em}\; \text{F} S \text{D} = \Z.\\
\end{aligned}
\end{equation}


\smallskip

We reformulate \eqref{lcnmf} by choosing the metric induced by the Frobenius norm as the distance measure and by translating the constraints in terms of distance. Considering then its Lagrangian form, we get the following unconstrained loss function to be minimized, with constraints incorporated as penalization terms:
\begin{equation}
\label{pennmf}
    \mathcal{L}(C, S) = 
    \lVert \text{X} - C S \rVert_{F}^2 +
    \alpha \lVert \text{B} C \text{A} - \text{Y} \rVert_{F}^2 +
    \beta \lVert \F S \D - \Z \rVert_{F}^2
\end{equation}
for some $\alpha, \beta > 0$ penalizing violations of the constraints.

To minimise $\mathcal{L}$, we start from arbitrary initial values and update the factor matrices as follows, until convergence:
$$S \leftarrow S \cdot \frac{C^\prime \X + \beta \F^\prime \Z \D^\prime}{C^\prime C S + \beta \F^\prime \F S \D \D^\prime}\, , \quad C \leftarrow C \cdot \frac{\X S^\prime+\alpha \B^\prime \Y \A^\prime}{CSS^\prime + \alpha \B^\prime \B C \A\A^\prime}$$
where $\cdot$ (multiplication) and $\frac{\phantom{AB}}{\phantom{DC}}$ (division) operations are coefficient-wise.
These updates are derived following the same logic as in \cite{lee_algorithms_2000}, further detailed in the appendix.

\smallskip

A first consideration regards the determination of suitable $\alpha$ and $\beta$ values as they define the weight that will be given to the terms penalizing the violation of the constraints. Given that the penalization loss terms may not be of the same magnitude as the factorization loss term, their primary purpose is a rescaling one. Analysing jointly the three loss terms at convergence obtained for different values of $(\alpha, \beta)$, one can determine suitable values.

A second consideration concerns the initialization strategy, which is obviously a critical point. Our approach involves running the MU algorithm $N=1000$ times, each time with a different initialization -- keeping the same initial value $\begin{bmatrix} \frac{1}{p} & \hdots & \frac{1}{p} \end{bmatrix}^\prime$ for the rows of $S$ but obtaining each time a different initial value for $C$ by uniformly sampling its rows on the simplex $\mathcal{S}(K) = \{ x \in \mathbb{R}_+^K \mid x_1 + \dots + x_K = 1 \}$ --
generating a large set of solutions. This set of solutions is then clustered based on the loss function value at convergence and the cluster of solutions associated with minimum value is retained. This allows us to work with an ensemble of $N^*$ solutions that can be combined to obtain more robust estimates and to quantify uncertainty. The ensemble estimated consumption $\hat{e}_{ik}$ of source $k$ during day $i$ is hence deduced by averaging the individual estimates given by the $N^*$ solutions:
$$\hat{e}_{ik} = \frac{1}{N^*} \sum_{l=1}^{N^*} \hat{c}_{ik}^{(l)} e_i$$
Additionally one can also retrieve source $k$'s estimated load during day $i$:
$$\hat{L}_{ik} = \frac{1}{N^*} \sum_{l=1}^{N^*} \hat{c}_{ik}^{(l)} e_i\, S_{k}^{(l)}$$
As mentioned in subsection \ref{subsec:quali}, we can project a new load curve $\X_{0}$ on the linear space spanned by the solution matrix $\hat{S}$ to estimate the sources concentrations hence sectors' contributions to $\X_{0}$. The source separation problem becomes ``non-blind" as we only need to estimate the concentrations:
$$\hat{C}_0=\argmin_{C \in \mathbb{R}_+^{n \times K}} \, \lVert \X_0 - C \hat{S} \rVert_{F}^2$$
This is a convex optimisation problem  for which the global minimum can be found, as long as the sources are linearly independent (i.e. $\hat{S}$ is full rank). In practice, we use the standard MU of \cite{lee_algorithms_2000} to solve the problem, updating $C$ only.

\section{Results}
\label{sec:res}

\subsection{LCNMF solutions analysis and interpretation}

We applied the source separation algorithm to 2021 and 2022 data,
acting as a training set, to identify the sources and consequently the sectors' profiles. We hold 2023 data out as a test set on which we estimated the concentrations of the sources identified in the training set, without incorporating the ASD-MSI information.
We then compared the sectors' monthly consumption of 2023 estimated from the decomposition with the ASD-MSI target.

\begin{figure}[h]
    \centering
    \includegraphics[width=0.6\linewidth]{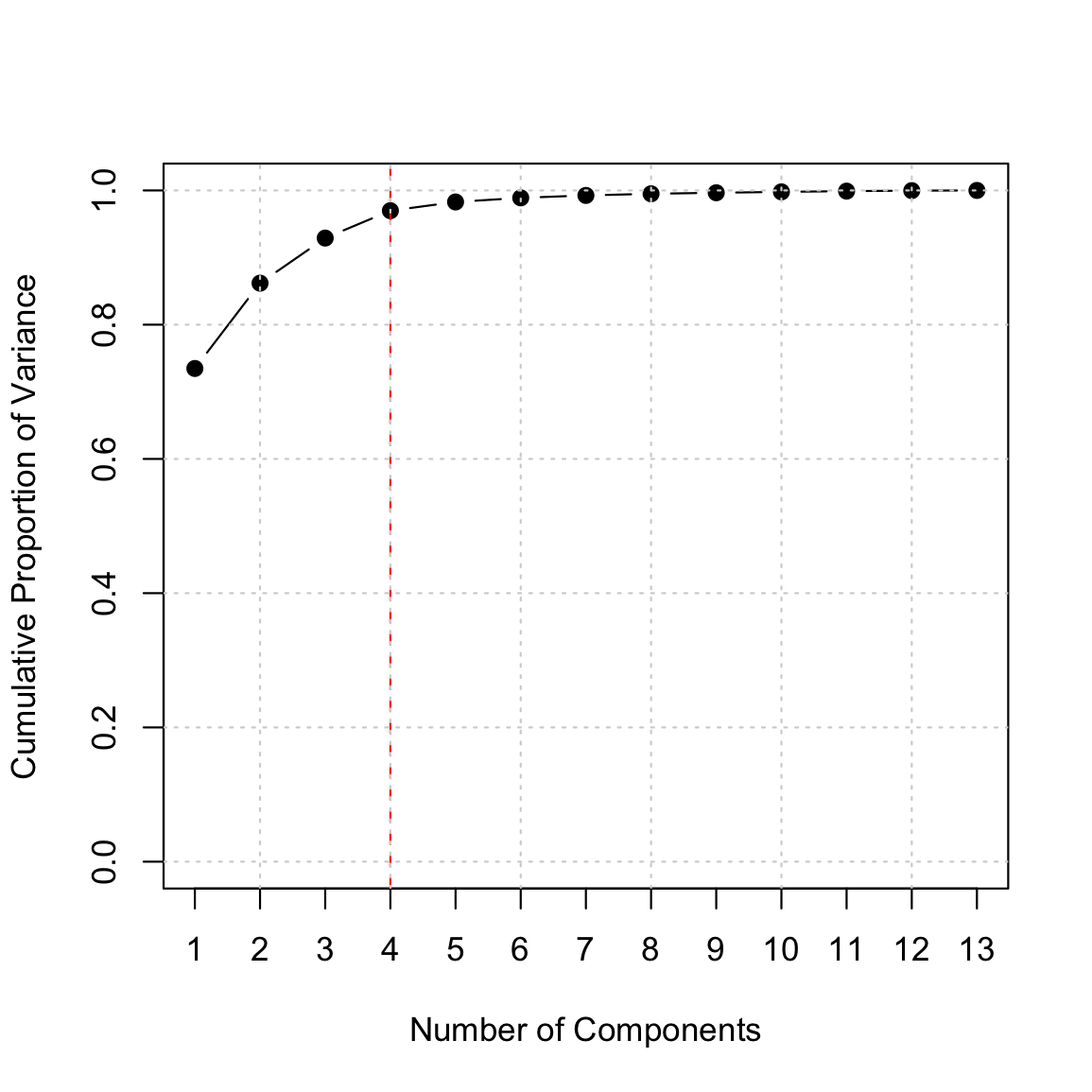}
    \caption{Scree plot of the FPCA ran on 2021-2022 load curves}
    \label{fig:scree}
\end{figure}

Fig. \ref{fig:scree} shows the scree plot of the FPCA performed on 2021-2022 load curves. We can see that 4 principal components explain $\approx 97\%$ of the variance and appear as a good choice for dimensionality reduction. We can therefore deduce that we should get a satisfying decomposition with $K=5$ sources.
Trials clearly indicate that the optimal sources/sectors mapping is obtained by assigning two sources to households, two sources to services and one source to industry.

\begin{figure}[h]
    \centering
    \includegraphics[width=0.8\linewidth]{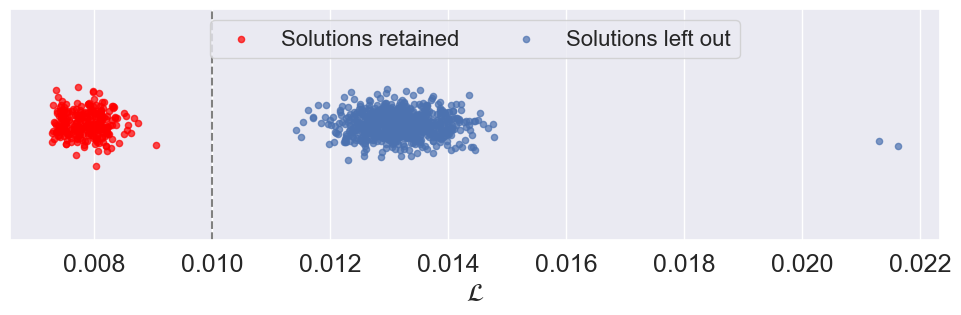}
    \caption{Value of the loss $\mathcal{L}$ at convergence for the $N=1000$ LCNMF runs. The values are vertically jittered for legibility. The grey dashed line indicates the threshold used to isolate the red cluster.
    }
    \label{fig:loss}
\end{figure}

Concerning the penalization parameters, we chose $\alpha=3\cdot10^{-10}$ and $\beta=1$ as they were providing the best trade-off between matrix factorization quality and compliance with the constraints. Following the $N=1000$ LCNMF runs, two compact and well-separated clusters of loss values at convergence were observed. We then simply used a threshold of $\mathcal{L}=0.01$ to retrieve the cluster of $N^*=271$ optimal solutions (Fig. \ref{fig:loss}).

As underlined in subsection \ref{subsec:sol}, we do not choose a specific solution but we analyse and use the entire set of $N^*$ optimal solutions to get a robust ensemble estimation and quantify uncertainty.


A first consideration regards the quality of the approximation of the load curves that we get with the LCNMF decomposition. Table \ref{tab:approx} compares the average low-rank matrix approximation quality of LCNMF with the approximation given by the projection of the curves on the space spanned by the first 4 functional principal components.

\begin{table}[h]
\centering
\begin{tabular}{lccc}
\toprule
\textit{Loss} & $\Vert A - B \Vert_1$ & $\Vert A - B \Vert_F$ & $\Vert A - B \Vert_\infty$ \\
\midrule
FPCA  & 9.2 & 0.0096 & 0.0053 \\
LCNMF & \textbf{8.9} & \textbf{0.0077} & \textbf{0.0051} \\
\bottomrule
\end{tabular}
\caption{Comparison of the low-rank matrix approximation quality between FPCA and LCNMF\label{tab:approx}}
\end{table}

The comparison is relevant because both approximation methods rely on a 4-dimensional representation. We observe that LCNMF provides a better fit than FPCA, showing that the approximation quality is more than satisfactory.

\smallskip


\begin{figure*}[!t]
\centering
\subfloat[$S_1$ (Household)]{\includegraphics[width=0.33\linewidth]{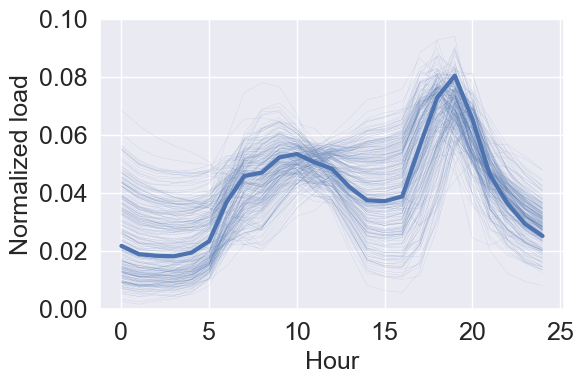}%
\label{s1}}
\subfloat[$S_2$ (Household)]{\includegraphics[width=0.33\linewidth]{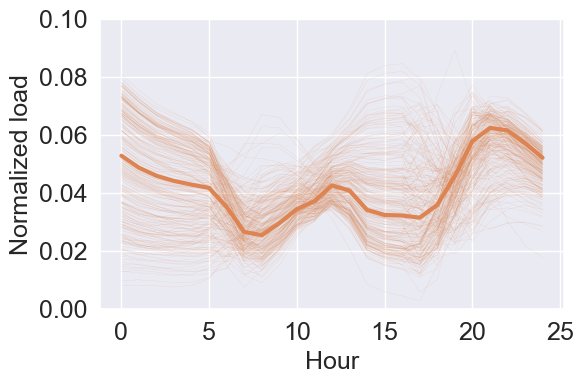}%
\label{s2}}
\subfloat[$S_3$ (Industry)]{\includegraphics[width=0.33\linewidth]{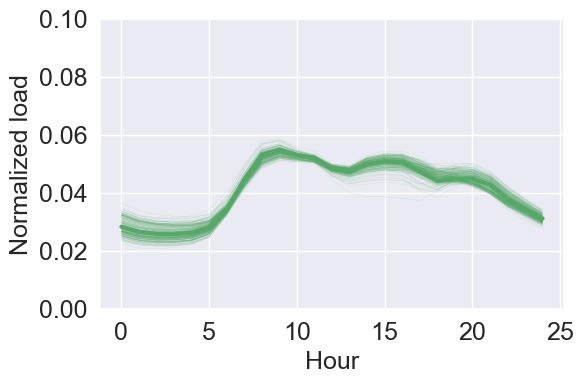}%
\label{s3}}
\hfil
\subfloat[$S_4$ (Services)]{\includegraphics[width=0.33\linewidth]{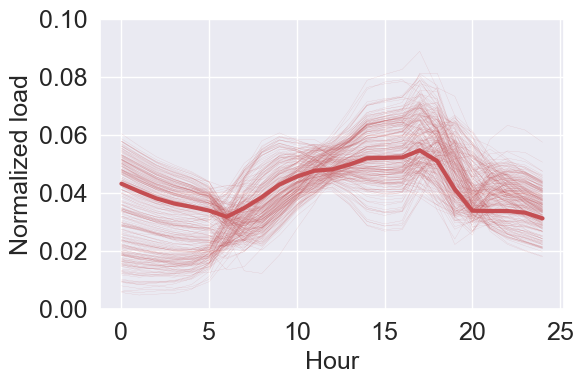}%
\label{s4}}
\subfloat[$S_5$ (Services)]{\includegraphics[width=0.33\linewidth]{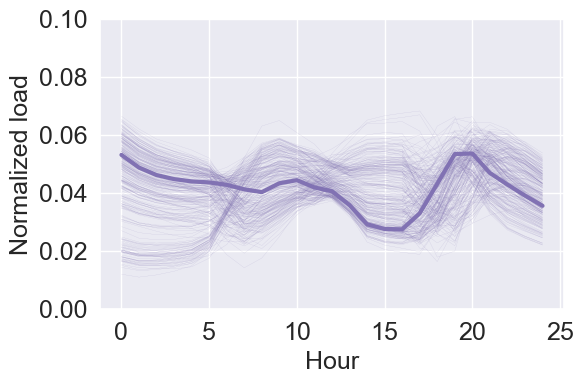}%
\label{s5}}
\caption{Sources corresponding to the $N^*=271$ retained solutions. Each thin line corresponds to one solution while the thick line, added for visualization purposes, is the geometric medoid of the thin lines, which is the closest observation to the geometric median (in terms of Euclidean distance). Since inside a sector the source number is arbitrary, the sources of household and services were aligned on the order given by the first solution.}
\label{fig:sources}
\end{figure*}

%

The extracted sources are shown in Fig. \ref{fig:sources}. We notice considerable variability across the different solutions, except for the source corresponding to the industrial sector. In this case, all solutions converge on the same load profile, marked by a sharp morning ramp beginning at 05:00 and three distinct peaks at 09:00, 16:00, and 20:00. Sources mapped to the households sector show a morning ramp that either starts steeply at 05:00 or more gradually at 08:00. The first one displays an evening peak around 18:00-19:00, while the second one shows it around 21:00-22:00. Beyond these few observations, the results are hard to interpret. The sources linked to the services sector are generally flatter than those of the household sector. They typically exhibit a morning ramp either at 05:00 or 07:00, depending on the solution. In particular, for most solutions, $S_4$ (Services) shows an afternoon peak around 17:00–18:00. $S_5$ (Services) seems to be characterized, in most cases, by a rather constant regime with an afternoon trough centered on 15:00 and an evening peak located around 20:00.

\begin{figure}[!t]
\centering
\subfloat[Households (per day type)]{\includegraphics[width=0.4\linewidth]{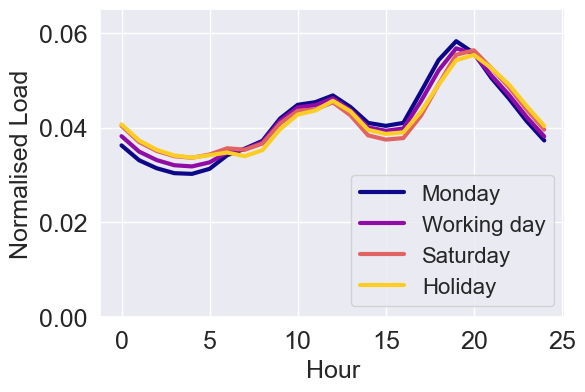}%
\label{fig:prof:dom:d}}
\subfloat[Households (per season)]{\includegraphics[width=0.4\linewidth]{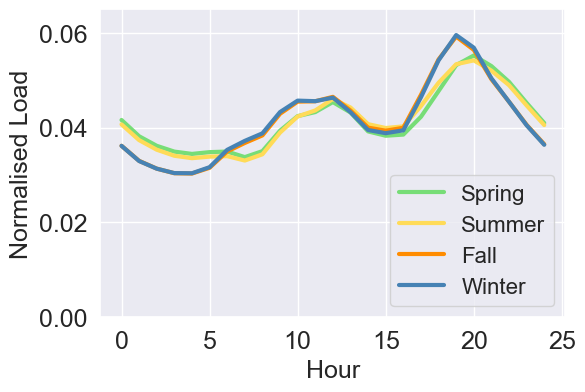}%
\label{fig:prof:dom:s}}
\hfil
\subfloat[Services (per day type)]{\includegraphics[width=0.4\linewidth]{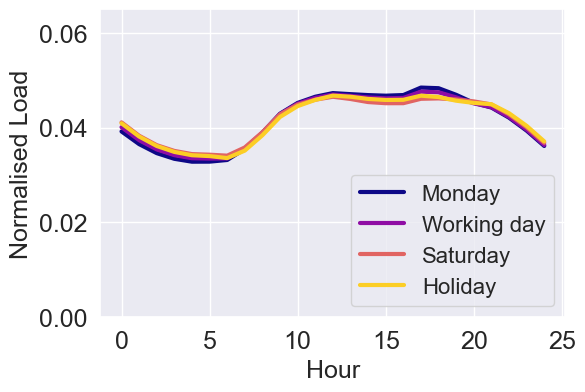}%
\label{fig:prof:ser:d}}
\subfloat[Services (per season)]{\includegraphics[width=0.4\linewidth]{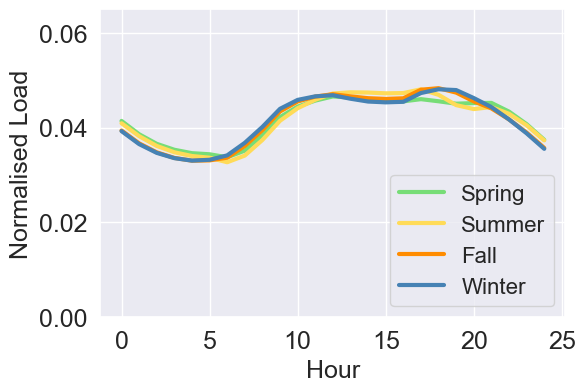}%
\label{fig:prof:ser:s}}
\caption{Average load profiles found for the households and services sectors depending on the day type and the season. The load profiles were first averaged at the day type and season level for each solution separately after which the average across solutions was retrieved. For legibility, we do not show the result for all individual solutions.}
\label{fig:prof}
\end{figure}

Overall, except for the industrial sector whose load profile is directly described by $S_3$ (Industry), the interpretability of the resulting sources for the two other sectors is limited as their load profile is given by a positive linear combination of their sources.
For these two sectors, there is in fact a profile specific to each day as the weights of the linear combination are free to vary, causing therefore a modification of the profile whenever one weight increases in proportion to the other. Fig. \ref{fig:prof} shows the average profile of households and services that we get depending on the type of day -- Monday, Working day, Saturday and Holiday\footnote{Sundays and Bank holidays} -- and the season\footnote{The mapping is made at the month level so winter is composed of all days from January to March, spring from April to June, summer from July to September and fall from October to December.}. It is quite clear that enabling flexible profiles allows us to model weekly and yearly seasonal variations in these two sectors' daily consumption patterns.
Specifically, we can see that the households profile is represented by a double-hump curve with a first small hump at 10:00-11:00 and a second higher around 19:00-20:00.
The curve is flatter and has a shallow and late morning ramp during the weekends compared to the weekdays. We can also note that the evening peak is delayed of one hour, occurring more around 20:00 than 19:00. Concerning the season effect, we observe a fall-winter vs a spring-summer regime. Fall-winter profiles show higher amplitude, earlier and steeper morning ramp, and earlier evening peak around 19:00. 
On average, the services profiles show a nearly constant activity between 10:00 and 20:00 with no clear morning or evening peak, which is consistent with businesses and offices hours in Italy. While the day type effect is not so clear, we observe a season effect that is similar to that for households, with the difference that summer and spring show slightly diverging behaviors in the afternoon.


\smallskip

The interpretation of the identified sources is made easier by looking at their concentrations measured every day. Specifically, we can try to understand if a source is more or less active depending on the day type and the period of the year.

\begin{figure}[!h]
\centering
\subfloat[Per day type]{\includegraphics[width=0.8\linewidth]{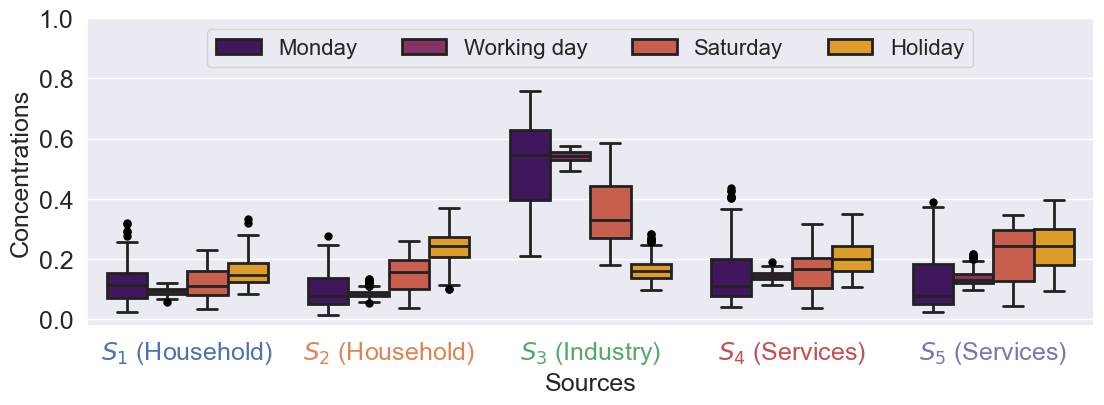}%
\label{fig:conc:daytype}}
\hfil
\subfloat[Per season]{\includegraphics[width=0.8\linewidth]{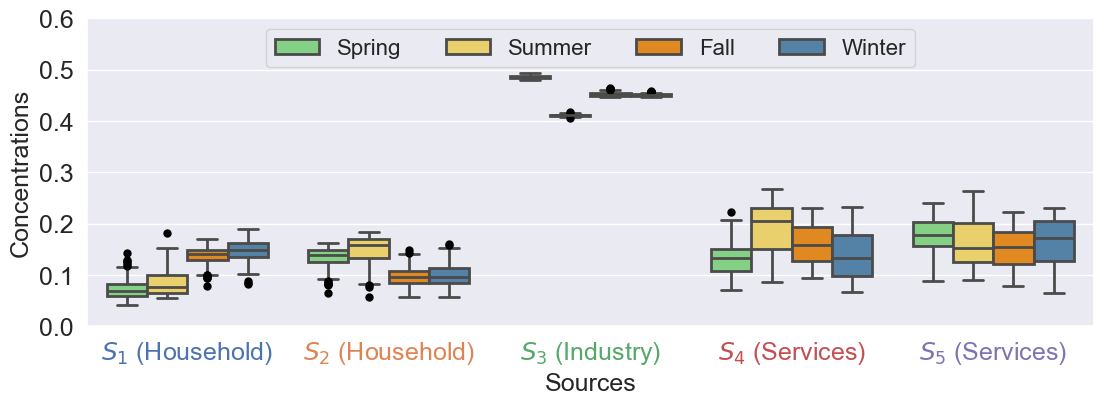}%
\label{fig:conc:season}}
\caption{Distribution of the average concentrations per day type and season across solutions.}
\label{fig:conc}
\end{figure}

Fig. \ref{fig:conc} shows the distribution of the average concentration of each source per type of day and season. We can therefore appraise both the effect of day type/season, and the variability, or disagreement, among the solutions.
Overall, the day type seems to have a clear effect on the activity of any source (fig. \ref{fig:conc:daytype}): on average, all five sources contribute equally to holidays while Mondays and working days are clearly dominated by $S_3$ (Industry). Saturdays lie in-between. Besides, we can note the strong agreement between solutions -- which can be called ``models" at this point -- on the average concentration estimated during working days compared to the other day types. On the contrary we observe considerable disagreement in the disaggregation of Mondays' load curves, especially for what regards $S_3$'s contribution.

For what regards the season effect, fig. \ref{fig:conc:season} highlights a clear opposition spring-summer vs fall-winter conditions for sources associated to households, $S_1$ and $S_2$, as was observed previously in the profiles. This observation supports the existence of two load regimes characterizing household consumption, one under cold temperatures, driven by $S_1$, and one under warm temperatures, driven by $S_2$. A possible explanation is the quadratic effect of temperature on power consumption in Italy \cite{pagliarini_outdoor_2019} with electric heating in winter and air conditioning devices in summer, causing two distinct seasonal shifts in the daily consumption profile. Note this dual effect may not apply to countries where there are just either heating or cooling needs, e.g., Denmark. The season effect is also pronounced on $S_3$ (Industry), with a lower activity in summer (due to factory closures in August) and a highest activity in spring, indicating therefore that this effect is more a calendar than a climatic one. A summer effect opposed to the spring one is also perceived for $S_4$ (Services).

\subsection{Disaggregated consumption}

The core output of the process, as indicated in Fig. \ref{fig:data-wf}, is the disaggregated load signal, with the three separate contributions of the sectors. We first compute the daily load curve of each source, obtained from the shape curve rescaled with the fraction of the daily consumption corresponding to that source, itself obtained by multiplying the concentration with the daily consumption. We then add up the sources curves to get the sectors curves, that we finally concatenate along days to get a continuous signal. Note that the signal presents an inherent discontinuity every day at midnight since the concentration varies non-smoothly from day to day.
This discontinuity is obviously a mathematical artifact, which however does not affect much the aggregated daily quantities.
Nonetheless, for applications where this discontinuity is deemed to be problematic, it can be overcome by smoothing the concentrations regarded as hourly data over time, keeping daily average concentrations equal to the original ones. In this paper, we don't pursue the matter further.

\begin{figure}[h!]
\centering
\subfloat[Aggregated load signal]{\includegraphics[width=0.7\linewidth]{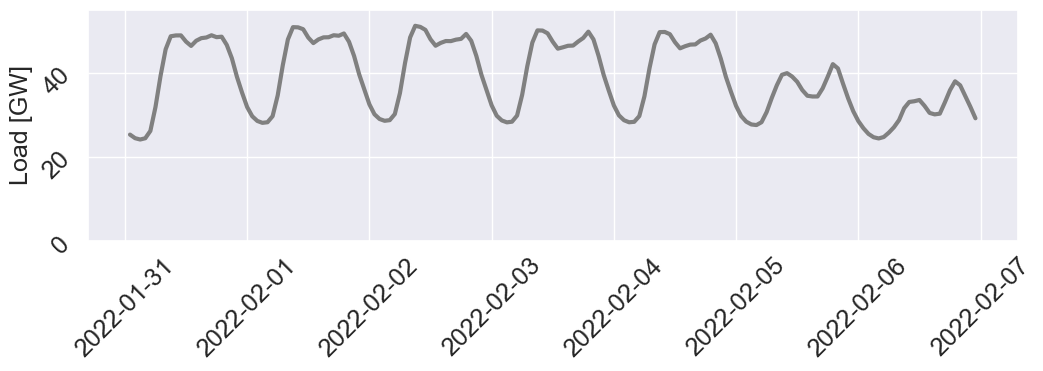}%
\label{fig:load:agg}}
\hfil
\subfloat[Disaggregated load signal]{\includegraphics[width=0.7\linewidth]{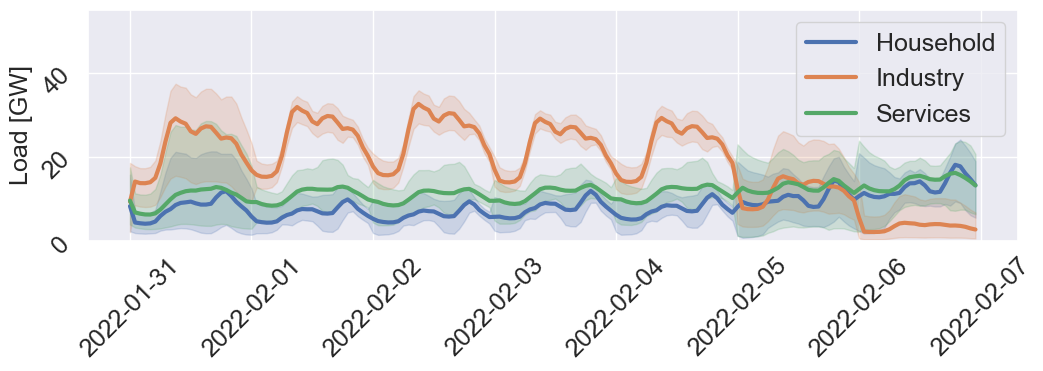}%
\label{fig:load:dis}}
\caption{Result of the load disaggregation for the week starting on Monday January 31\textsuperscript{st}, 2022. Shaded areas on the bottom panel represent the one-at-a-time 95\% prediction intervals given by the 0.025 and 0.975 quantiles of the ensemble estimation.}
\label{fig:load}
\end{figure}

The disaggregated load signal for the first week of February 2022 is shown on Fig. \ref{fig:load:dis}. Beyond the uncertainty that is quite high for the household and services sectors compared to the industry one, we can observe that the uncertainty is higher during the weekend and even higher on Monday. This confirms the earlier observed fact that Mondays have a different status from the other weekdays.

Unfortunately, we don't have sector-level consumption data to validate the results of the disaggregation at the hourly, daily or weekly level. But moving to the monthly level, we have the MSI to which our estimates can be compared, which are the data, together with the ASD, that were used to guide the separation process. As 2023 was left out in the blind separation process -- meaning that the sector-level information corresponding to that year is completely unknown to the model -- we can reliably validate our disaggregation procedure by verifying whether monthly sectors' consumption estimates align with the MSI for that year. \rev{As a benchmark, we include a naive forecast corresponding to one-year-lagged values and a yearly-seasonal exponential smoothing model: Holt-Winter's (HW) additive method with no trend\footnote{Given the series are stationary.} \citep{hyndman_forecasting_2021}. For the HW model, we forecast the 12 months of 2023 using 2021-2022 data only -- similarly to the BSS approach.}


\begin{figure}[!h]
\centering
\subfloat[Household]{\includegraphics[width=0.33\linewidth]{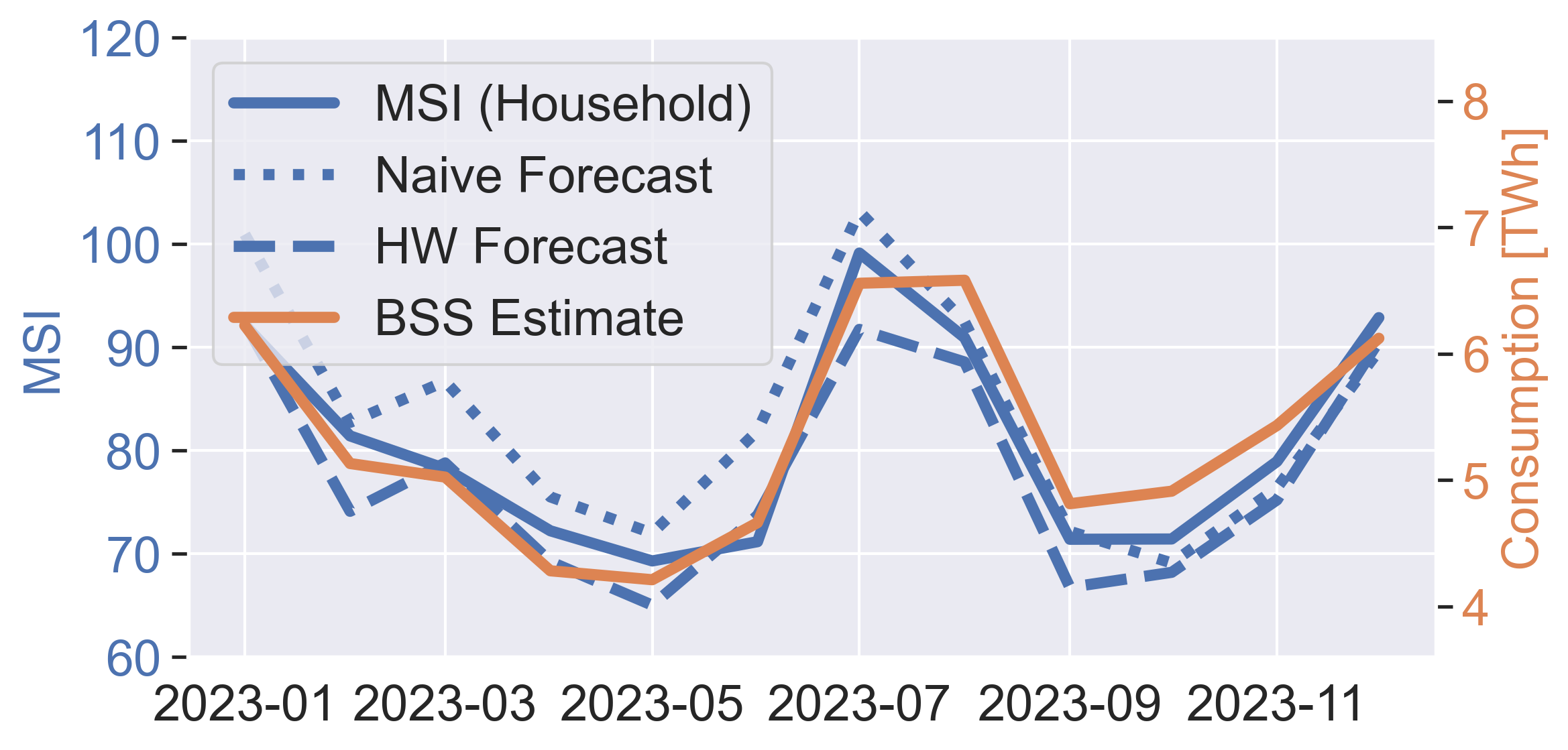}%
\label{fig:month:dom}}
\hfil
\subfloat[Industry]{\includegraphics[width=0.33\linewidth]{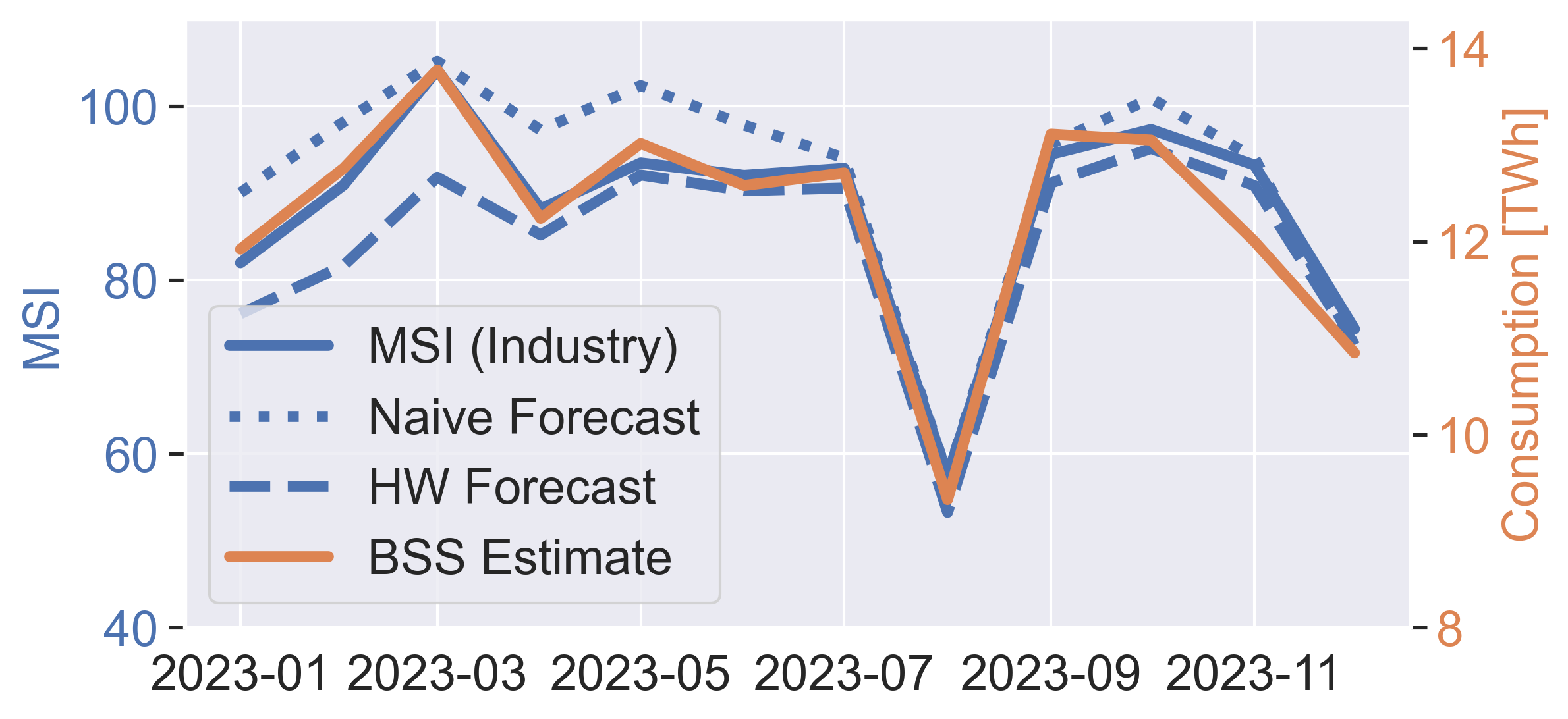}%
\label{fig:month:ind}}
\hfil
\subfloat[Services]{\includegraphics[width=0.33\linewidth]{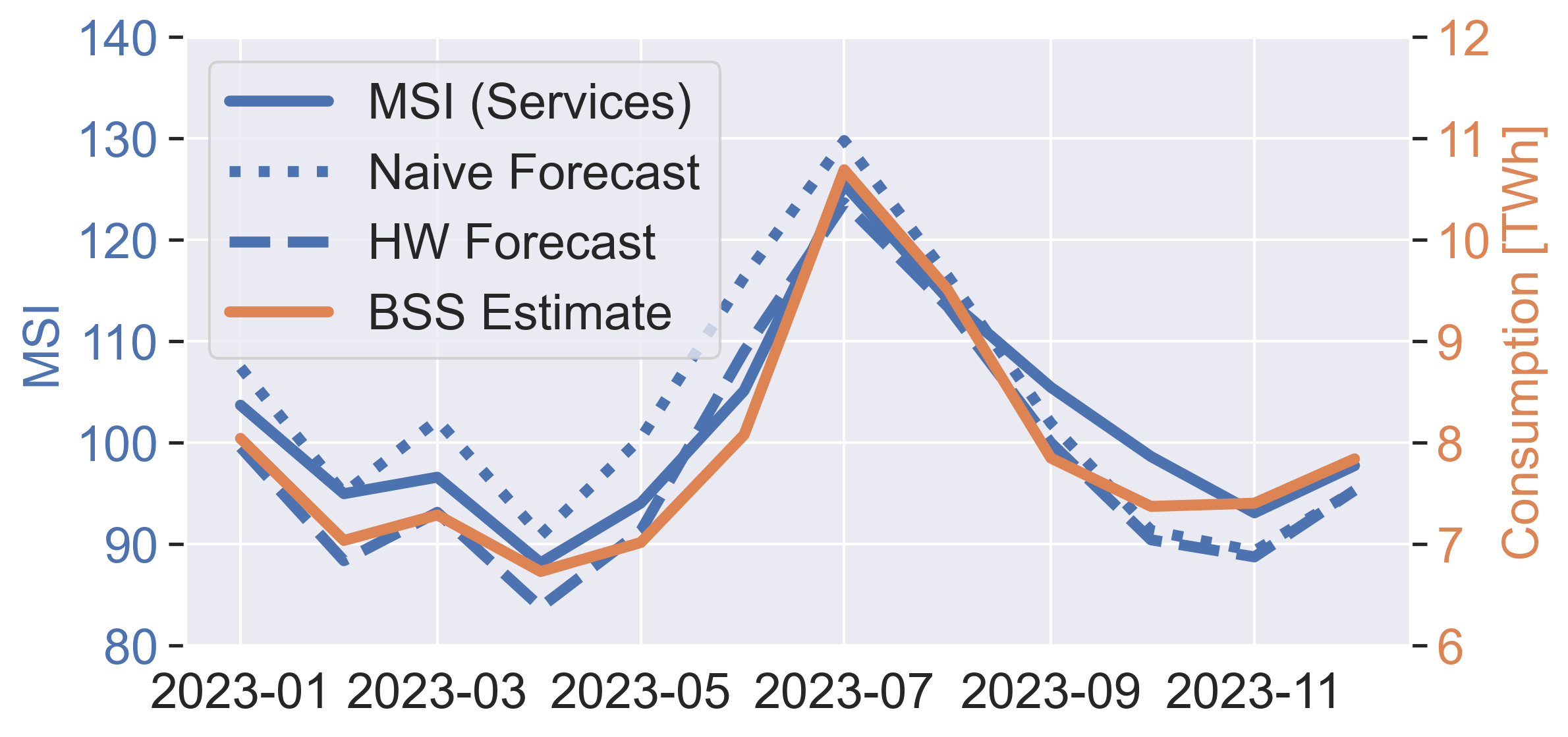}%
\label{fig:month:ser}}
\caption{Comparison of the BSS monthly sector consumption estimates with the TSO's MSI and \rev{its naive and HW forecasts} for the test year 2023. Note that the data are not on the same scale and are therefore represented with two different ordinate axis (left/blue for MSI and right/orange for BSS).}
\label{fig:month}
\end{figure}

\begin{table}[h]
\centering
\begin{tabular}{rccl}
\toprule
 & Naive forecast & \rev{HW forecast} & BSS \\
\midrule
Household  & 0.914 & \rev{0.959} & 0.951 \\
Industry & 0.962 & \rev{0.961} & 0.974 \\
Services & 0.912 & \rev{0.971} & 0.970 \\
\bottomrule
\end{tabular}
\caption{Pearson's correlation coefficient with MSI for the test year 2023.}
\label{tab:corr}
\end{table}

The comparison of these out-of-sample predictions with the MSI for 2023 is reported in Fig. \ref{fig:month} \rev{and Table \ref{tab:corr}}. \rev{All three estimation methods align well with the observed values. While BSS appears to reproduce the dynamics of the observed series slightly better than the naive forecast, its performance remains comparable to that of HW. These results suggest that the BSS approach produces accurate monthly consumption estimates and is competitive for nowcasting sectors monthly consumption.}

\rev{
\section{Discussion}
\label{sec:disc}
}
\rev{Though designed for the specific problem tackled in this work, The proposed LCNMF method is highly generalizable as it allows to solve a load disaggregation problem at any scale of the power network, for which prior information -- such as low-frequency disaggregated consumption values or direct information on one or more load components -- can be expressed as linear constraints on the factor matrices.  While the current formulation addresses linear constraints only, the framework can be easily adapted to support more complex non-linear forms, provided they involve functions differentiable with respect to the factor matrices (see \ref{app1}).}

\medskip

\rev{The application of our BSS framework to Italian national load data revealed established features of household, industry, and services sectors' electricity consumption profiles.}
\rev{There is in fact a strong consistency between the profiles we estimate and those identified in previous studies relying on ground truth data \cite{andersen_longterm_2013, gerossier_novel_2017, voulis_understanding_2018}, thereby supporting the validity of our results and the effectiveness of the proposed method.
In particular, households consumption shows a daily profile that is dominated by an evening peak and is driven by two distinct regimes, in warmer vs colder seasons. The services sector is found to be especially active during the usual businesses and office hours and shows a singular summer daily load profile, which could be related to the air conditioning devices working at their maximum around 16:00, and only in summer. These patterns globally align with those observed in \cite{andersen_longterm_2013} for Denmark, with some nuances likely due to climate and lifestyle differences. Finally, industrial consumption is particularly low during holiday periods and does not seem to be much affected by seasons. The trimodal pattern we identify for the industrial daily load profile is very similar to that reported by \cite{andersen_longterm_2013} and \cite{gerossier_novel_2017}. Besides, the fact that a single source satisfactorily describes industry's daily load profile, unlike household and services profiles which are described by linear combinations of two sources each, is in line with the findings of \cite{andersen_longterm_2013} on the invariance of industrial power consumption. This consistent daily consumption pattern allows for a clear distinction of industry from the other two sectors. In contrast, households and services profiles appear to show a certain degree of similarity that makes them harder to dissociate.}
\rev{Besides, the monthly sectoral consumption estimates derived from disaggregated load were found to closely align with the TSO's official statistics -- indices based on consumer samples -- and be a competitive approach for nowcasting sectors' monthly consumption.}

\medskip

\rev{A primary limitation of this application is the lack of ground truth data at fine temporal resolutions, such as hourly or daily, which would enable (i) improved supervision of the source separation process and (ii) a direct assessment of disaggregation accuracy, thereby complementing the literature-based validation.}
\rev{Besides, the model assumes a hierarchical relationship between the identified sources and the sectors, an assumption that may not hold in practice. Indeed, a given source may be associated with multiple sectors -- for example, air conditioning usage in both residential and commercial consumption. Finally, we do not explicitly control for weather effects which are likely to introduce additional variability in the load signal. Properly accounting for such effects, by incorporating weather data directly in the model,  may yield more consistent load profiles and lower estimation uncertainty. Another possible solution is to adopt a methodology to separate the weather-dependent load component, such as that of \cite{dampeyrou_unsupervised_2024}, prior to the sectoral separation.}
\rev{Future research could extend the empirical validation to other countries or smaller power zones where high-frequency ground truth consumption data are available, potentially considering a more granular consumer segmentation.}

\bigskip

\rev{From a broader perspective, models capable of detecting and profiling consumption patterns, distinguishing among different categories of customers -- such as that proposed in the present work, are a valuable tool for flexibility management. Indeed, while generation is becoming more diffused and less predictable, power demand is evolving due to new drivers such as electric vehicles, heat pumps and data centers. Concurrently, business opportunities emerge, resulting in a much more complex energy market structure. Within this framework, system operators may lack the resources to monitor all the assets necessary to effectively manage the power system. Consequently, they could benefit from tools able to reconstruct sector-aware power profiles with appropriate granularity. This possibly unlocks opportunities to efficiently schedule balancing actions, such as shifting, curtailment and modulation of loads, thereby reducing peak generation and the need for costly reserves. Moreover, the use of advanced, standardized profiling models for end-user demand can enhance TSO-DSO coordination, improving system-wide balancing and power system resilience.}

\medskip

\rev{Another central takeaway from this work is the potential of statistical decomposition methods for solving electrical load profiling or disaggregation problems \textit{beyond the building scale}, even when no disaggregated consumption measurements are available at high-frequency. We hope this contribution will encourage further exploration of this topic.}

\rev{\section{Conclusion}}
\label{sec:conc}

Monitoring electricity consumption at the level of residential, commercial and industrial sectors is of considerable interest to the TSO and other power system stakeholders. To date, no solution exists to perform it live, at country scale, high frequency and utilizing open-access national grid load measurements only. In this context, we introduced a novel signal separation framework to monitor sectors consumption from \rev{open-access} aggregate grid-scale load measurements collected by the TSO. \rev{The proposed framework is highly generalizable as it allows to solve a load disaggregation problem at any scale of the power network where disaggregated ground truth data are not available, provided there is some prior information on load components which can be expressed as linear constraints on the factor matrices.}
\rev{Applied to Italian national load data over the 2021-2023 period, our method identifies residential, commercial and industrial daily load profiles which are consistent with existing literature. Besides, sectoral monthly consumption estimates obtained from load disaggregation align with official sector-level power demand statistics used by the TSO.}

\rev{The main limitation of this study lies in the absence of quantitative validation at hourly or daily level, due to the lack of ground truth data at these levels. The application to other countries or smaller zones where granular disaggregated load data are available is left for future research.}

\smallskip


\bigskip

\section*{Code and data availability}
\noindent The analyses presented in this work were conducted using Python 3.11 and R 4.3.0. Code and data are available at \url{https://github.com/gkchln/load-bss}. Note that the monthly sector indicator \cite{terna_enel_2024} cannot be shared for privacy reasons and is therefore not included in the repository.

\section*{Acknowledgments}
\noindent The authors want to thank Valeria Amoretti and Luca Marchisio of Terna S.p.A for their contribution to the interesting and deep discussions on the topics developed in this paper.

\section*{Credit author statement}
\noindent \textbf{Guillaume Koechlin:} Conceptualization, Methodology, Software, Writing - Original Draft, Visualization. \textbf{Filippo Bovera:} Conceptualization, Methodology, Writing - Review \& Editing. \textbf{Elena Degli Innocenti:} Conceptualization, Methodology, Resources, Validation, Writing - Review \& Editing. \textbf{Barbara Santini:} Conceptualization, Resources, Validation, Writing - Review \& Editing. \textbf{Alessandro Venturi:} Conceptualization, Resources, Supervision, Validation, Writing - Review \& Editing. \textbf{Simona Vazio:} Conceptualization, Resources, Validation, Writing - Review \& Editing. \textbf{Piercesare Secchi:} Conceptualization, Methodology, Writing - Review \& Editing.

\section*{Funding}
\noindent The present research has been partially supported by MUR, grant Dipartimento di Eccellenza 2023-2027.

\section*{Declaration of generative AI and AI-assisted technologies in the writing process}
\noindent During the preparation of this work the authors used ChatGPT in order to obtain suggestions for improving writing fluency and clarity. After using this tool/service, the authors reviewed and edited the content as needed and take full responsibility for the content of the published article.

\clearpage
\appendix
\section{Modified multiplicative updates for the resolution of LCNMF}
\label{app1}

\noindent
Given $\B \in \mathbb{R}_+^{m \times n}$, $\A \in \mathbb{R}_+^{K \times g}$, $\Y \in \mathbb{R}_+^{m \times g}$, $\E \in \mathbb{R}_+^{l \times K}$, $\D \in \mathbb{R}_+^{p \times q}$, $\Z \in \mathbb{R}_+^{l \times q}$,
we propose to solve:
\begin{equation}
\label{lcnmf_app}
    (\hat{C},\hat{S})=\argmin_{C \in \mathbb{R}_+^{n \times K},\, S \in \mathbb{R}_+^{K \times p}}  \mathcal{L}(C, S, \alpha, \beta)
\end{equation}
where
$$\mathcal{L}(C, S, \alpha, \beta) = \lVert \text{X} - C S \rVert_{F}^2 +
    \alpha \lVert \text{B} C \text{A} - \text{Y} \rVert_{F}^2 +
    \beta \lVert \text{E} S \text{D} - \text{Z} \rVert_{F}^2$$
and $\alpha, \beta$ are two non-negative penalization parameters.

\medskip

\noindent The case $\alpha=\beta=0$ corresponds to the ordinary NMF problem:
$$\mathcal{L}(C, S) = \mathcal{L}(C, S, 0, 0)= \lVert \text{X} - C S \rVert_{F}^2$$
Using matrix differential calculus, with $\phantom{}^\prime$ denoting the transpose:
$$\nabla_S\mathcal{L}(C, S) = 2C^\prime CS - 2C^\prime \text{X} \quad \text{and} \quad \nabla_C\mathcal{L}(C, S) = 2CSS^\prime - 2\text{X}S^\prime$$
Separating positive and negative contributions, the gradient $\nabla_S \mathcal{L}$ or $\nabla_C \mathcal{L}$ can be written:
$$\nabla_S\mathcal{L}(C, S) = \nabla_S^+\mathcal{L}(C, S) - \nabla_S^-\mathcal{L}(C, S)$$

The popular multiplicative updates (MU) algorithm of Lee and Seung \cite{lee_algorithms_2000}, in case of Frobenius loss, proposes to alternatively update $C$ and $S$, as follows:
$$S \leftarrow S \cdot \frac{C^\prime \text{X}}{C^\prime C S}\, , \quad C \leftarrow C \cdot \frac{\text{X}S^\prime}{CSS^\prime}$$
Where $\cdot$ and $\frac{\phantom{AB}}{\phantom{DC}}$ operations are coefficient-wise.

There are several ways to derive these updates, mentioned in the original paper of Lee and Seung and detailed in section 8.2 of \cite{gillis_nonnegative_2020}, among which the ``gradient ratio heuristic" that shows that these updates can be written:
$$C \leftarrow S \cdot \frac{\nabla_S^-\mathcal{L}(C, S)}{\nabla_S^+\mathcal{L}(C, S)}\, , \quad C \leftarrow C \cdot \frac{\nabla_C^-\mathcal{L}(C, S)}{\nabla_C^+\mathcal{L}(C, S)}$$

\medskip

\noindent We can apply the same idea in the case $\alpha \neq 0$ or $\beta \neq 0$. In this case we have:
$$\nabla_C\mathcal{L}(C, S, \alpha, \beta) = 2CSS^\prime + 2\alpha \B^\prime \B C \A \A^\prime - (2\X S^\prime + 2\alpha \B ^\prime \Y \A^\prime)$$
$$\nabla_S\mathcal{L}(C, S, \alpha, \beta) = 2C^\prime CS + 2\beta \E^\prime \E S \D \D^\prime - (2C^\prime \X + 2\beta \E^\prime \Z \D^\prime)$$
and we can therefore consider the following updates:
$$S \leftarrow S \cdot \frac{C^\prime \X + \beta \E^\prime \Z \D^\prime}{C^\prime C S + \beta \E^\prime \E S \D \D^\prime}\, , \quad C \leftarrow C \cdot \frac{\X S^\prime+\alpha \B^\prime \Y \A^\prime}{CSS^\prime + \alpha \B^\prime \B C \A\A^\prime}$$
With a well-chosen auxiliary function, the proofs in \cite{lee_algorithms_2000} can be directly adapted to prove that the previous updates lead to a monotonically decreasing loss $\mathcal{L}$ and a converging update sequence.



\bibliographystyle{elsarticle-num-names}
\bibliography{references}



\end{document}